\documentclass[authoryear]{FLO_v1}

\usepackage[utf8]{inputenc}			
\usepackage[T1]{fontenc}			
\usepackage[british]{babel}	
\usepackage{amsmath}				
\usepackage{amssymb}				
\usepackage{mathtools}				
\usepackage{bm}						
\usepackage{graphicx}				
\usepackage{xcolor} 				
\usepackage{comment} 				
\usepackage{array}					
\usepackage{tabularx}				
\usepackage[english=british]{csquotes} 
\usepackage[shortlabels]{enumitem} 

\usepackage{ragged2e}

\usepackage{upgreek}
\usepackage{multicol,multirow}
\usepackage{amsmath,amssymb,amsfonts}
\usepackage{mathrsfs}
\usepackage{amsthm}
\usepackage[figuresright]{rotating}
\usepackage{appendix}
\usepackage[authoryear]{natbib}
\usepackage{ifpdf}
\usepackage{newtxtext}
\usepackage{newtxmath}
\usepackage{textcomp}
\usepackage[colorlinks,allcolors=blue]{hyperref}
\definecolor{jourcolor}{cmyk}{1,0.57,0.01,0.38}
\hypersetup{
    colorlinks,%
    citecolor=jourcolor,%
    filecolor=jourcolor,%
    linkcolor=jourcolor,%
    urlcolor=jourcolor
}
\numberwithin{equation}{section}
\articletype{RESEARCH ARTICLE}
\DOI{TBA}
\Year{2024}
\Vol{X}
\Price{}
\art-id{FLOXXXX}
\citearticle{Vieweg, P. P. \& Käufer, T. \& Cierpka, C. \& Schumacher, J.}
\Enumber{Ex} 

\renewcommand{\Pr}		{\mathrm{Pr}} 

\newcommand{\Ra}   		{\mathrm{Ra}}

\newcommand{\Nu}   		{\mathrm{Nu}}

\newcommand{\Nuexp}   	{\mathrm{Nu}_{\textrm{exp}}} 
\renewcommand{\Re}		{\mathrm{Re}} 

\newcommand{\dT}   		{\Delta T}
\newcommand{\dTN}   	{\Delta T_{\textrm{N}}}

\newcommand{\Tinfty}   	{T_{\infty}} 
\newcommand{\Bi}   	    {\mathrm{Bi}}
\newcommand{\Tc}   	    {T_{\textrm{c}}} 
\newcommand{\Tt}   	    {T_{\textrm{t}}} 
\newcommand{\Tb}   	    {T_{\textrm{b}}} 
\newcommand{\Th}   	    {T_{\textrm{h}}} 

\newcommand{\tauf}   	{\tau_{\textrm{f}}} 

\begin{document}

\title[Digital twin of a large-aspect-ratio Rayleigh-Bénard experiment]{Digital twin of a large-aspect-ratio Rayleigh-Bénard experiment: Role of thermal boundary conditions, measurement errors and uncertainties}

\author{Philipp P. Vieweg$^{1\ast, 2}${\href{https://orcid.org/0000-0001-7628-9902}{\includegraphics{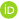}}}} 
\author{Theo Käufer$^{2}${\href{ https://orcid.org/0000-0001-7121-8291}{\includegraphics{orcid_logo}}}}
\author{Christian Cierpka$^{2}${\href{https://orcid.org/0000-0002-8464-5513}{\includegraphics{orcid_logo}}}}
\author{Jörg Schumacher$^{2, 3}${\href{https://orcid.org/0000-0002-1359-4536}{\includegraphics{orcid_logo}}}}

\address[1]{Department of Applied Mathematics and Theoretical Physics, Wilberforce Road, Cambridge, CB3 0WA, United Kingdom}
\address[2]{Institute of Thermodynamics and Fluid Mechanics, Technische Universität Ilmenau, Postfach 100565, 98684 Ilmenau, Germany.}
\address[3]{Tandon School of Engineering, New York University, New York, NY 11021, USA.}

\corres{*}{Corresponding author. E-mail:
\emaillink{ppv24@cam.ac.uk}}


\date{\textbf{Received:} 26 08 2024; \textbf{Revised:} 26 11 2024; \textbf{Accepted:} 29 11 2024}

\abstract{
Albeit laboratory experiments and numerical simulations have proven themselves successful in enhancing our understanding of long-living large-scale flow structures in horizontally extended Rayleigh-Bénard convection, some discrepancies with respect to their size and induced heat transfer remain. This study traces these discrepancies back to their origins.
We start by generating a digital twin of one standard experimental set-up. This twin is subsequently simplified in steps to understand the effect of non-ideal thermal boundary conditions, and the experimental measurement procedure is mimicked using numerical data.
Although this allows explaining the increased observed size of the flow structures in the experiment relative to past numerical simulations, our data suggests that the vertical velocity magnitude has been underestimated in the experiments. A subsequent re-assessment of the latter's original data reveals an incorrect calibration model. The re-processed data show a relative increase in $u_{z}$ of roughly $24 \%$, resolving the previously observed discrepancies.
This digital twin of a laboratory experiment for thermal convection at Rayleigh numbers $\Ra = \left\{ 2, 4, 7 \right\} \times 10^{5}$, a Prandtl number $\Pr = 7.1$, and an aspect ratio $\Gamma = 25$ highlights the role of different thermal boundary conditions as well as a reliable calibration and measurement procedure.
}

\maketitle

\begin{boxtext}
\textbf{\mathversion{bold}Impact Statement}
The formation and dynamics of large-scale flow structures in horizontally extended turbulent Rayleigh-Bénard convection is essential for an understanding of its heat transfer. By creating a digital twin of a laboratory experiment, we investigate the influence of realistic thermal boundary conditions, which always deviate from ideal ones, and measurement deviations on the discrepancies between experimental and simulation results. The insights gained have broad implications for engineering and technological heat transfer applications. Understanding these effects can improve thermal management systems in industrial processes and electronic devices and provide critical guidance for future laboratory setups in fluid mechanics studies.
\end{boxtext}

\section{Introduction}
\label{sec:Introduction}
The presence of convection as one of the basic means of heat transfer is of paramount importance for many natural systems -- including habitable conditions on Earth -- and engineering problems.
Understanding it allows, for instance, to predict (space) weather \citep{Atkinson1996, Schwenn2006, Pulkkinen2007}, to exploit the induced pressure gradients across Earth's atmosphere by wind turbines \citep{Hau2013, Vallis2017}, or even to power electrical devices which's thermal management has been optimised to get rid of an active fan \citep{Shabany2010}.

\begin{figure}
\centering
\includegraphics[scale = 1.0]{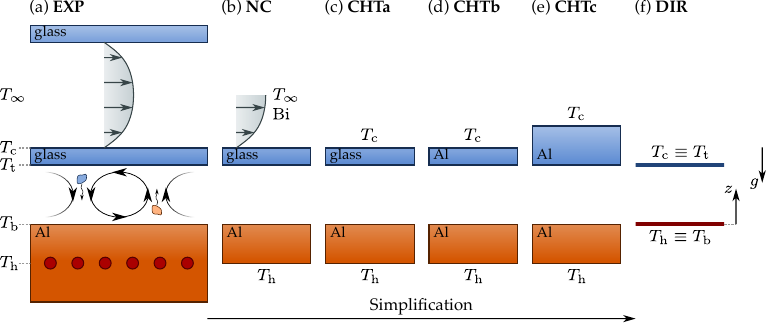}
\caption{\justifying{Schematic configurations.
We take the motivating laboratory experiment (EXP) on the left, create its digital twin which involves a Newton cooling (NC) condition, and subsequently simplify the latter successively. Identifiers for different configurations are included above each sketch. The location of different temperatures is defined on the left, whereas panels (b -- f) include the corresponding control parameters only; other values manifest dynamically. CHT stands for regular conjugate heat transfer and DIR for pure or classical Dirichlet boundary conditions.}}
\label{fig:schematic_configurations}
\end{figure}

This broad applicability of general insights on naturally-driven thermal convection has been attracting an uncountable number of researchers over the past century towards its paradigm, Rayleigh-Bénard convection. 
Whilst the key idea of the latter is to transfer heat through a horizontal layer of fluid of thickness $H$ that is heated from below and cooled from above while being subjected to gravity, the particular set-up can differ significantly depending on the approach. 
Figure \ref{fig:schematic_configurations} depicts in panels (a) and (f) typical configurations present in laboratory experiments \citep{Moller2020, Moller2022} and numerical simulations \citep{Vieweg2021, Vieweg2023a}, respectively. 
Undoubtedly, our progress is partly driven by both the symbiosis and antibiosis between these different approaches, exploiting either complex measurement techniques or expensive computing facilities to generate data.

Arrangements like these have allowed to prove the existence of \textit{long-living large-scale flow structures} \citep{Vieweg2023, Vieweg2023a, Vieweg2024a} in horizontally extended domains despite being superposed to turbulence on significantly smaller time and length scales. 
Depending on the thermal boundary conditions applied at the heated bottom and cooled top plane, those roll-like flow structures (see figure \ref{fig:schematic_configurations} (a)) exhibit different properties. In a nutshell, one observes either \textit{turbulent superstructures} \citep{Pandey2018, Vieweg2021} with a characteristic horizontal extension of $\Lambda_{\textrm{char}} \sim \mathcal{O} \left( H \right)$ or the gradual aggregation of convection cells towards a flat convection roll or \textit{supergranule} \citep{Vieweg2021, Vieweg2022b, Vieweg2023a, Vieweg2024, Vieweg2024a} with $\Lambda_{\textrm{char}} \gg \mathcal{O} \left( H \right)$ depending on whether the temperature field or its vertical gradient, respectively, is spatially homogeneous at the horizontal boundaries of the fluid. These two situations are physically linked to limits of the ratio of thermophysical properties between the fluid and its adjacent solid \citep{Hurle1967, Chapman1980, Chapman1980a, Otero2002}. 
One way to quantify this ratio is via the thermal diffusivity $\kappa$ \citep{Hurle1967} -- the latter of which controls the (time-dependent) relaxation of thermal perturbations occurring at the solid-fluid interface --, such that these different forcings correspond to $\kappa_{\textrm{st, sb}} / \kappa_{\textrm{fl}} \rightarrow \infty$ and $\kappa_{\textrm{st, sb}} / \kappa_{\textrm{fl}} \rightarrow 0$, respectively. Interestingly, the thermal boundary conditions seem to dominate any variation of the strength of the thermal driving (as quantified via the Rayleigh number $\Ra$) or working fluid (as specified by the Prandtl number $\Pr$) in $3$-dimensional analyses \citep{Vieweg2021, Vieweg2023a, Vieweg2024a}.

Of course, these limits can only be approximated by real materials and material selection is often further constrained by requirements of the measurement techniques such as optical transparency. Figure \ref{fig:schematic_configurations} (a) shows such a compromise when analysing the large-scale heat transfer patterns \citep{Moller2020, Moller2022}. On the one hand, the glass plate on top of the fluid layer of interest allows optical access, but it leads to $\kappa_{\textrm{st}} / \kappa_{\textrm{fl}} = 2.5$ and needs to be cooled using a pressure-driven flow. On the other hand, the aluminium bottom plate offers $\kappa_{\textrm{sb}} / \kappa_{\textrm{fl}} = 435$ and can be heated quite uniformly via meandering channels. Despite a small temperature difference of approximately $2 K$ between both plates \citep{Moller2022a, Moller2022}, a convection flow emerges in the water which is visualised by suspended (temperature-sensitive) tracer particles in a horizontal slab at mid-height. A detailed description of the laboratory experiment can be found in \cite{Moller2022}. 
This asymmetry of thermal boundary conditions has so far not been accounted for in simulations, resulting in unresolved discrepancies when comparing experimental and numerical results.
Most strikingly, experiments suggested an \textit{in}creased size of the turbulent superstructures and \textit{de}creased induced heat transfer across the fluid layer compared to simulations \citep{Pandey2018, Moller2021, Moller2022a, Vieweg2021, Vieweg2021a, Vieweg2022, Vieweg2023}. 
Unfortunately, precise attributions of those disagreements have not yet been possible due to a lack of simulations that resemble the experiment -- especially with respect to its thermal boundary conditions, strength of thermal driving, and horizontal extent of the domain; all of which affect those disagreements -- sufficiently. 

This study aims to resolve the disagreements between laboratory experiments and numerical simulations by creating a digital twin which mimics the former's geometrical and thermophysical properties as well as boundary conditions.
First, we study the impact of non-ideal and asymmetric thermal boundary conditions by an iterative simplification of this initial numerical configuration towards the classical numerical set-up (i.e., without solid plates). Although this exposes strong thermal variations at the upper solid-fluid interface for the digital twin and allows to explain an increased size of flow structures, the observed heat transfer disagrees even stronger. 
Thus and second, we successively modify the twin's true numerical data to imitate the experimentally present measurement procedure. This includes a systematic spatial averaging over interrogation windows, an erroneous detection of the mean solid-fluid interface temperatures, and uncertainties for the particle image thermometry.
Contrasting the modified numerical with experimental data at $\Ra = \left\{ 2, 4, 7 \right\} \times 10^{5}$, we find strong disagreements for the vertical or out-of-plane component of the velocity field. 
A subsequent re-assessment of the original experimental data confirms that an incorrect calibration model led in the past to a systematic underestimation of this velocity component and consequently also to a reduced perceived heat transfer.
Hence, this study highlights how digital twins of laboratory experiments can help aligning the results of experimental and numerical approaches and understanding their discrepancies in detail.

\section{Numerical method}
\label{sec:Numerical_method}

\subsection{Governing equations}
\label{subsec:Governing_equations}
Given the tiny mean temperature difference across the fluid layer in the motivating experimental configuration -- see again figure \ref{fig:schematic_configurations} (a) --, we consider an incompressible flow based on the Oberbeck-Boussinesq approximation \citep{Oberbeck1879, Boussinesq1903}. This means that material parameters are assumed to be constant except for the mass density, the latter of which varies at first order (with respect to temperature) only when acting together with gravity \citep{Rayleigh1916, Vieweg2023a}. 

The three-dimensional equations of motion are solved by the spectral-element method Nek5000 \citep{Fischer1997,Scheel2013}. We non-dimensionalise the equations based on characteristic quantities of the fluid domain such as the fluid layer height $H$ and temperatures at the bottom and top of this fluid layer, $\Tb$ and $\Tt$, respectively. The characteristic (dimensional) temperature scale $\dT := \langle \Tb - \Tt \rangle_{A, t}$ is based on the mean temperatures across the corresponding horizontal cross-section $A$ and time $t$. Together with the free-fall inertial balance, the free-fall velocity $U_{\textrm{f}} = \sqrt{\alpha g \dT H}$ and time scale $\tau_{\textrm{f}} = H / U_{\textrm{f}} = \sqrt{H / \alpha g \dT}$ establish as further characteristic units. The pressure scale is $p_{\textrm{f}} = U_{\textrm{f}}^{2} \rho_{\textrm{ref, fl}}$. Here, $\alpha$ is the volumetric thermal expansion coefficient of the fluid at constant pressure, $g$ the acceleration due to gravity, and $\rho_{\textrm{ref, fl}}$ the reference density of the fluid at reference temperature. 

Despite this general approach, the governing equations differ depending on the phase of the domain. 
The equations relevant to the fluid domain translate into
\begin{align}
\label{eq:CE}
\nabla \cdot \bm{u} &= 0 , \\
\label{eq:NSE}
\frac{\partial \bm{u}}{\partial t} + \left( \bm{u} \cdot \nabla \right) \bm{u} &= - \nabla p + \sqrt{\frac{\Pr}{\Ra}} \thickspace \nabla^{2} \bm{u} + T \bm{e}_{z} , \\
\label{eq:EE_fl}
\frac{\partial T}{\partial t} + \left( \bm{u} \cdot \nabla \right) T &= \frac{1}{\sqrt{\Ra \Pr}} \thickspace \nabla^{2} T.
\end{align}
For the solid domains, one obtains a pure diffusion equation,
\begin{align}
\label{eq:EE_s}
\frac{\partial T}{\partial t} &= \frac{\kappa_{\Phi}}{\kappa_{\textrm{fl}}} \frac{1}{\sqrt{\Ra \Pr}} \thickspace \nabla^{2} T, 
\end{align}
since the velocity is zero therein. In any of these equations, $\bm{u}$, $T$ and $p$ represent the (non-dimensional) velocity, temperature and pressure field, respectively. 

The relative strength of the individual terms in the \textit{fluid}-related equations \eqref{eq:CE} -- \eqref{eq:EE_fl} is specified by nothing but the Rayleigh and Prandtl number, 
\begin{equation}
\label{eq:def_Rayleigh_number_Prandtl_number}
\Ra = \frac{\alpha g \dT H^{3}}{\nu \kappa} \qquad \textrm{and} \qquad \Pr = \frac{\nu}{\kappa} .
\end{equation}
The quantities $\nu$ and $\kappa$ denote the kinematic viscosity and thermal diffusivity of the fluid, respectively.
In contrast, the \textit{solid}-related equation \eqref{eq:EE_s} requires the additional specification of the thermal diffusivity $\kappa_{\Phi} \equiv \lambda_{\textrm{t, } \Phi} / \rho_{\Phi} c_{p \textrm{, } \Phi}$ of a solid domain of interest $\Phi = \left\{ \textrm{st}, \textrm{sb} \right\}$ relative to $\kappa_{\textrm{fl}} \equiv \lambda_{\textrm{t, fl}} / \rho_{\textrm{ref, fl}} c_{p \textrm{, fl}}$ from the fluid domain -- hence, the ratio of thermal diffusivities $\kappa_{\Phi} / \kappa_{\textrm{fl}}$ turns up as additional control parameter.
Here, $\lambda_{\textrm{t}}$ represents the thermal conductivity, $\rho_{\Phi}$ the mass density of the solid domain, and $c_{p}$ the specific heat capacity at constant pressure. 
This coefficient results simply from the above non-dimensionalisation based on parameters of the fluid domain together with the definitions in equation \eqref{eq:def_Rayleigh_number_Prandtl_number}. In this work, we use the subscripts $\left\{ \textrm{fl}, \textrm{st}, \textrm{sb} \right\}$ to indicate the fluid, solid top, or solid bottom domain, respectively.

\subsection{Numerical domain and its boundary conditions}
\label{subsec:Numerical_domain_and_its_boundary_conditions}
Resembling the laboratory experiment, the governing equations \eqref{eq:CE} -- \eqref{eq:EE_s} are complemented by a closed three-dimensional domain with a square horizontal cross-section $A = \Gamma \times \Gamma$ and an aspect ratio $\Gamma := L / H$ where $L$ is the horizontal length of the domain. The thickness of the solid top and bottom domains is defined via their respective aspect ratio $\Gamma_{\Phi} := H_{\Phi} / H$ and varies (just like their thermophysical properties) between the different configurations depicted in figure \ref{fig:schematic_configurations}. The solid bottom domain is thus situated at $- \Gamma_{\textrm{sb}} \leq z \leq 0$, the fluid domain at $0 \leq z \leq 1$, and the solid top domain at $1 \leq z \leq 1 + \Gamma_{\textrm{sb}}$.

The fluid obeys at any of its boundaries no-slip boundary conditions, 
\begin{equation}
\label{eq:BC_no_slip}
\bm{u} = 0 \qquad \textrm{at all fluid boundaries.}
\end{equation}
Moreover, we assume perfectly thermally insulated lateral boundaries such that
\begin{equation}
\label{eq:BC_insulation}
\frac{\partial T}{\partial x} \left( x = \pm \Gamma / 2 \right) = \frac{\partial T}{\partial y} \left( y = \pm \Gamma / 2 \right) = 0 .
\end{equation}
The thermal boundary conditions at the different horizontal boundaries can conceptually be classified into (i) internal or passive and (ii) external or active ones. 
Concerning the former, the two (potential) solid-fluid interfaces require the continuity of both the temperature field and diffusive heat flux ($\bm{J}_{\textrm{dif}} = - \lambda_{\textrm{t}} \nabla T$ in the dimensional framework), i.e., 
\begin{subequations}
\label{eq:BC_interface}
\begin{alignat}{6}
\label{eq:BC_interface_bot}
T_{\textrm{sb}} &= T_{\textrm{fl}} \quad &
&\textrm{and} \quad &
&\frac{\kappa_{\textrm{sb}}}{\kappa_{\textrm{fl}}} &&\frac{1}{\sqrt{\Ra \Pr}} &\frac{\partial T_{\textrm{sb}}}{\partial z} &= \frac{1}{\sqrt{\Ra \Pr}} \frac{\partial T_{\textrm{fl}}}{\partial z} &
\qquad \textrm{at } z &= 0 , \textrm{ as well as} \\
\label{eq:BC_interface_top}
T_{\textrm{st}} &= T_{\textrm{fl}} \quad &
&\textrm{and} \quad &
&\frac{\kappa_{\textrm{st}}}{\kappa_{\textrm{fl}}} &&\frac{1}{\sqrt{\Ra \Pr}} &\frac{\partial T_{\textrm{st}}}{\partial z} &= \frac{1}{\sqrt{\Ra \Pr}} \frac{\partial T_{\textrm{fl}}}{\partial z} &
\qquad \textrm{at } z &= 1 .
\end{alignat}
\end{subequations}
We term the corresponding temperature fields at these interfaces $\Tb := T \left( z = 0 \right)$ and $\Tt := T \left( z = 1 \right)$. The precise spatio-temporal temperature and heat flux distributions at these two boundaries manifest dynamically depending on the fluid flow, the latter of which is induced by the external thermal boundary conditions at $z = \left\{ - \Gamma_{\textrm{sb}}, 1 + \Gamma_{\textrm{st}} \right\}$ in the first place. 
In the laboratory experiment, see again figure \ref{fig:schematic_configurations} (a), these conditions differ significantly. 
We thus apply a classical Dirichlet as well as a Newton cooling boundary condition ($\lambda_{\textrm{t}} \nabla T \cdot \bm{n} = h_{\textrm{conv}} \left( T - \Tinfty \right)$ with the dimensional convection coefficient $h_{\textrm{conv}}$ and wall-normal unity vector $\bm{n}$ in the dimensional framework) to its digital twin, such that
\begin{subequations}
\label{eq:BC_bot_top}
\begin{alignat}{2}
\label{eq:BC_Dirichlet_bot}
T \left( z = - \Gamma_{\textrm{sb}} \right) &= \Th 
\qquad &&\textrm{and} \\
\label{eq:BC_Newton_cooling}
\frac{\kappa_{\textrm{st}}}{\kappa_{\textrm{fl}}} \frac{1}{\sqrt{\Ra \Pr}} \frac{\partial T}{\partial z} &= \Bi \thickspace \frac{\kappa_{\textrm{st}}}{\kappa_{\textrm{fl}}} \frac{1}{\sqrt{\Ra \Pr}} \left( T - \Tinfty \right)
\qquad &&\textrm{at } z = 1 + \Gamma_{\textrm{st}} 
\end{alignat}
\end{subequations}
at the bottom and top of the numerical domain, respectively. 
Note that the nature of these boundary conditions is quite different: while the former fixes the temperature itself at the boundary, the latter couples the local vertical heat flux to the temperature difference between the present temperature field and the undisturbed temperature $\Tinfty$ of the convectively cooling fluid. Hence, the Newton cooling boundary condition is less strict and allows for a non-uniform temperature distribution. 
Furthermore, it requires the additional quantification of the strength of convective cooling relative to thermal conduction at the corresponding boundary via the Biot number 
\begin{equation}
\label{eq:def_Biot_number}
\Bi = \frac{h_{\textrm{conv}} H}{\lambda_{\textrm{t, st}}} .
\end{equation}

In the iterative process of simplifying the configuration, see again figure \ref{fig:schematic_configurations}, we might eventually substitute this Newton cooling (NC) boundary condition from equation \eqref{eq:BC_Newton_cooling} by another Dirichlet condition 
\begin{equation}
\label{eq:BC_Dirichlet_top}
T \left( z = 1 + \Gamma_{\textrm{st}} \right) = \Tc .
\end{equation}
At the end of this process, the solid domains will be omitted entirely and the internal thermal boundary conditions \eqref{eq:BC_interface} will disappear -- this is the case in most numerical Rayleigh-Bénard convection studies \citep{Pandey2018, Krug2020, Vieweg2021} and depicted in figure \ref{fig:schematic_configurations} (f).

Let us briefly summarise the different non-dimensional parameters that have been introduced. 
In a nutshell, we have collected (i) control parameters from the governing equations itself $\left\{ \Ra, \Pr \right\}$, (ii) geometric parameters $\left\{ \Gamma, \Gamma_{\textrm{sb}}, \Gamma_{\textrm{st}} \right\}$, (iii) thermophysical parameters $\left\{ \kappa_{\textrm{sb}} / \kappa_{\textrm{fl}}, \kappa_{\textrm{st}} / \kappa_{\textrm{fl}} \right\}$, as well as (iv) thermal boundary condition parameters which are either $\left\{ \Th, \Bi, \Tinfty \right\}$ or $\left\{ \Th, \Tc \right\}$ depending on the precise configuration. Thereby, we have also introduced the temperature fields $\left\{ \Tinfty, T_{\textrm{c}}, T_{\textrm{t}}, T_{\textrm{b}}, T_{\textrm{h}} \right\}$ at the coordinates shown in figure \ref{fig:schematic_configurations} (a). Note that the internal temperatures at the solid-fluid interfaces are a function of space and time, whereas the external temperatures will be considered to be constant.

\subsection{Initial condition}
\label{subsec:Initial_condition}
The applied external temperatures can be used to compute a one-dimensional, stationary diffusive temperature profile across the different layers given their geometrical and thermophysical properties. Altered by some tiny random thermal noise $0 \leq \Upsilon \leq 10^{-3}$ and together with a fluid at rest (i.e., $\bm{u} \left( t = 0 \right) = 0$), this profile is used as initial condition for each simulation.

The stationary temperature profiles within the 3 layers at hand can generally be expressed via 
\begin{alignat}{7}
\label{eq:IC_st}
&T_{\textrm{st}} &&\left( z_{\textrm{st}}, t = 0 \right) &&= T_{\textrm{t}} - \left( T_{\textrm{t}} - T_{\textrm{c}} \right) &&\frac{z_{\textrm{st}}}{\Gamma_{\textrm{st}}} \qquad &&\textrm{for } z_{\textrm{st}} &&\equiv z - 1 &&\in \left[ 0, \Gamma_{\textrm{st}} \right] , \\
\label{eq:IC_fl}
&T_{\textrm{fl}} &&\left( z_{\textrm{fl}}, t = 0 \right) &&= T_{\textrm{b}} - \left( T_{\textrm{b}} - T_{\textrm{t}} \right) && \medspace z_{\textrm{fl}} \qquad &&\textrm{for } z_{\textrm{fl}} &&\equiv z &&\in \left[ 0, 1 \right] , \\
\label{eq:IC_sb}
&T_{\textrm{sb}} &&\left( z_{\textrm{sb}}, t = 0 \right) &&= T_{\textrm{h}} - \left( T_{\textrm{h}} - T_{\textrm{b}} \right) &&\frac{z_{\textrm{sb}}}{\Gamma_{\textrm{sb}}} \qquad &&\textrm{for } z_{\textrm{sb}} &&\equiv z + \Gamma_{\textrm{sb}} &&\in \left[ 0, \Gamma_{\textrm{sb}} \right] .
\end{alignat}
Since the diffusive heat flux needs to match at the various interfaces, these profiles are coupled. In the case of a present Newton cooling (i.e., given $\left\{ \Th, \Bi, \Tinfty \right\}$), the boundary conditions from equation \eqref{eq:BC_bot_top} form together with equation \eqref{eq:BC_interface} a linear system of equations and yield
\begin{subequations}
\label{eq:IC_NewtonBC}
\begin{align}
\label{eq:IC_NewtonBC_Tc}
\Tc &= \left\{ \left[ h_{\textrm{st}} \left( h_{\textrm{fl}} + h_{\textrm{sb}} \right) + h_{\textrm{fl}} h_{\textrm{sb}} \right] h_{\Bi} \thickspace \Tinfty + h_{\textrm{st}} h_{\textrm{fl}} h_{\textrm{sb}} \thickspace \Th \right\} / D , \\
\label{eq:IC_NewtonBC_Tt}
\Tt &= \left\{ h_{\textrm{st}} \left( h_{\textrm{fl}} + h_{\textrm{sb}} \right) h_{\Bi} \thickspace \Tinfty + \left( h_{\textrm{st}} + h_{\Bi} \right) h_{\textrm{fl}} h_{\textrm{sb}} \thickspace \Th \right\} / D , \\
\label{eq:IC_NewtonBC_Tb}
\Tb &= \left\{ h_{\textrm{st}} h_{\textrm{fl}} h_{\Bi} \thickspace \Tinfty + \left[ h_{\textrm{st}} \left( h_{\textrm{fl}} + h_{\Bi} \right) + h_{\textrm{fl}} h_{\Bi} \right] h_{\textrm{sb}} \thickspace \Th \right\} / D 
\end{align}
with
\begin{equation}
\label{eq:IC_NewtonBC_D_hBi}
\qquad \quad 
D := h_{\textrm{st}} h_{\textrm{fl}} h_{\textrm{sb}} + \left[ h_{\textrm{st}} \left( h_{\textrm{fl}} + h_{\textrm{sb}} \right) + h_{\textrm{fl}} h_{\textrm{sb}} \right] h_{\Bi} , \quad 
h_{\Bi} := \Bi \thickspace \frac{\kappa_{\textrm{st}}}{\kappa_{\textrm{fl}}} \frac{1}{\sqrt{\Ra \Pr}} ,
\end{equation}
\vspace{-0.25cm}
\begin{equation}
\label{eq:IC_NewtonBC_h}
h_{\textrm{st}} := \frac{\kappa_{\textrm{st}}}{\kappa_{\textrm{fl}}} \frac{1}{\sqrt{\Ra \Pr}} \frac{1}{\Gamma_{\textrm{st}}} , \quad 
h_{\textrm{fl}} := \frac{\Nu}{\sqrt{\Ra \Pr}} , \quad 
h_{\textrm{sb}} := \frac{\kappa_{\textrm{sb}}}{\kappa_{\textrm{fl}}} \frac{1}{\sqrt{\Ra \Pr}} \frac{1}{\Gamma_{\textrm{sb}}} 
\end{equation}
\end{subequations}
where the last line describes the various heat transfer coefficients $h_{\Phi}$ or thermal conductances $h_{\Phi} \Gamma^{2}$ in the solid and fluid domains.
In the opposing case where the Newton condition \eqref{eq:BC_Newton_cooling} is substituted by the second Dirichlet condition \eqref{eq:BC_Dirichlet_top} (i.e., given $\left\{ \Th, \Tc \right\}$), 
\begin{subequations}
\label{eq:IC_DirichletBC}
\begin{align}
\label{eq:IC_DirichletBC_Tt}
\Tt &= \frac{\left( h_{\textrm{st}} + h_{\textrm{st}} h_{\textrm{sb}} / h_{\textrm{fl}} \right) \thickspace \Tc + h_{\textrm{sb}} \thickspace \Th}{h_{\textrm{st}} + h_{\textrm{sb}} + h_{\textrm{st}} h_{\textrm{sb}} / h_{\textrm{fl}}} , \\
\label{eq:IC_DirichletBC_Tb}
\Tb &= \Tt + \frac{h_{\textrm{st}}}{h_{\textrm{fl}}} \left( \Tt - \Tc \right)
\end{align}
\end{subequations}
can be deduced.

Remember that we used the temperature drop across the fluid layer as a characteristic quantity of the system in section \ref{subsec:Governing_equations}. However, as it might have become clear earlier from equation \eqref{eq:BC_interface} or here from the above initial conditions in equations \eqref{eq:IC_NewtonBC} and \eqref{eq:IC_DirichletBC}, the temperatures at these solid-fluid interfaces manifest dynamically. Considering the non-dimensionalisation of this mean temperature drop across the fluid layer, i.e., $\langle \Tb - \Tt \rangle_{A, t} \equiv \langle T \left( z = 0 \right) - T \left( z = H \right) \rangle_{A, t} = \dT \thickspace \langle \tilde{T} \left( z = 0 \right) - \tilde{T} \left( z = 1 \right) \rangle_{\tilde{A}, \tilde{t}} = \dT \thickspace \widetilde{\dT}$ \citep{Otero2002, Vieweg2023a} where tildes indicate non-dimensional quantities, this implies that one needs either to account for the non-dimensional temperature drop $\widetilde{\dT} \equiv \dTN$ in various equations or to make sure that $\dTN \simeq 1$ by adjusting the external temperatures correspondingly. We have decided to go with the latter option as this resembles the common situation in Rayleigh-Bénard convection.

Furthermore, note in particular that $h_{\textrm{fl}}$ from equation \eqref{eq:IC_NewtonBC_h} represents the \textit{effective} thermal conductance of the fluid layer as it comprises the Nusselt number $\Nu \geq 1$ (see equation \ref{eq:def_Nusselt_number}). Considering that the initial condition represents a fluid at rest, one might set $\Nu \left( t = 0 \right) = 1$. However, in this work we assume $\Nu \left( t = 0 \right) > 1$ to account for the thermal capacities of the different layers and aim reaching statistically steady conditions thereby more quickly. Of course, the final statistically stationary value of $\Nu$ is not known \textit{a priori}. 

Hence, we run in this work various preliminary 2- and 3-dimensional simulations in advance of each production simulation in order to find the optimal $\left\{ \Th, \Bi, \Tinfty \right\}$ or $\left\{ \Th, \Tc \right\}$ as well as the induced global heat transfer $\Nu$, the latter of which is then used in the initial condition. This iterative procedure ensures that $\dTN \simeq 1$ right from the initialisation.

\section{The impact of non-ideal thermal boundary conditions}
\label{sec:impact_of_non_ideal_thermal_boundary_conditions}

\subsection{The digital twin}
\label{subsec:The_digital_twin}
In order to study the effect of \textit{experimentally present} thermal boundary conditions, we start by creating a digital twin of the motivating laboratory experiment. The experiment configuration -- see again figure \ref{fig:schematic_configurations} (a) -- suggests to apply the Dirichlet and Newton cooling boundary conditions from equation \eqref{eq:BC_bot_top} at the bottom and top, respectively. The aspect ratio of the closed domain $\Gamma = 25$, the thickness of the aluminium bottom plate $\Gamma_{\textrm{sb}} = 0.66$ with a relative thermal diffusivity $\kappa_{\textrm{sb}} / \kappa_{\textrm{fl}} = 435$, and $\Gamma_{\textrm{st}} = 0.29$ with $\kappa_{\textrm{st}} / \kappa_{\textrm{fl}} = 2.5$ for the glass top plate. The resulting twin is sketched in figure \ref{fig:schematic_configurations} (b). Furthermore, we consider $\Ra = 2 \times 10^{5}$ and $\Pr = 7.1$ just like in the experiment \citep{Moller2022a, Moller2022}. 

Note that while thermal conduction through or in the lateral walls can become significant especially for slender convection cells \citep{Stevens2014}, we assume them to be adiabatic in our motivating experimental set-up (and thus also the digital twin, see eq. \eqref{eq:BC_insulation}) due to several reasons. First, the experiment offers a large aspect ratio $\Gamma \gg 1$ and so the total area of the side walls is by an order of magnitude smaller than that of the top and bottom plates. Second, the mean temperature in the turbulent flow is held at room temperature during experiments and the side walls are additionally insulated. And third, both the fluid's heat transfer coefficient and horizontal cross section are significantly larger than those of the side walls, rendering vertical heat transfer within the latter negligible.

The experimentally present convection coefficient $h_{\textrm{conv}}$ -- entering the non-dimensional control parameter $\Bi$ -- cannot be extracted from the collected experimental data. We thus estimate its value based on two different approaches. 
On the one hand, we consider a one-dimensional flow over a heated flat plate of uniform temperature with its evolving laminar boundary layer. On the other hand, we consider the Sieder-Tate law for fully transitioned laminar pipe flows based on the corresponding hydraulic diameter \citep{Incropera1996}. These two approaches yield $\Bi \simeq 5.1$ and $\Bi \simeq 6.0$, respectively, and we will consider both of these values in the following.

\subsection{Simplification of the numerical domain}
\label{subsec:Simplification_of_the_numerical_domain}
From the perspective of the convective fluid layer, the thermal boundary condition is determined by both the thermophysical and geometric properties of a solid plate adjacent to it. 
Firstly, if the ratio of thermal diffusivities $\kappa_{\Phi} / \kappa_{\textrm{fl}} \rightarrow \infty$, thermal perturbations relax much quicker in the solid plate compared to the fluid and so the former is an iso-thermal perfect thermal conductor. Vice versa, the plate appears as thermal insulator and the provided (constant) heat flux is independent of the fluid motion in the opposite case of $\kappa_{\Phi} / \kappa_{\textrm{fl}} \rightarrow 0$ \citep{Hurle1967}. Note further that the Nusselt number affects the effectively present ratio via $\kappa_{\textrm{fl, eff}} = \Nu \thickspace \kappa_{\textrm{fl}}$.
Secondly, also the geometry plays a crucial role. Consider therefore the horizontal and vertical thermal diffusion time scales $\tau_{\kappa \textrm{, } \Phi \textrm{, h}} := L_{\Phi}^{2} / \kappa_{\Phi}$ and $\tau_{\kappa \textrm{, } \Phi \textrm{, v}} := \left( \Gamma_{\Phi} H \right)^{2} / \kappa_{\Phi}$, respectively, inside an infinitely thin solid plate. In this case of $\Gamma_{\Phi} \rightarrow 0$, the external thermal boundary condition affects immediately also the solid-fluid interface since $\tau_{\kappa \textrm{, } \Phi \textrm{, v}} \ll \tau_{\kappa \textrm{, } \Phi \textrm{, h}}$. In other words, the external thermal boundary condition cannot be relaxed by thermal diffusion inside the solid plate and thus leaves a significant footprint on the solid-fluid interface. This footprint vanishes only in the opposite limit $\tau_{\kappa \textrm{, } \Phi \textrm{, v}} \gg \tau_{\kappa \textrm{, } \Phi \textrm{, h}}$, i.e., for $\Gamma_{\Phi} H \gg L_{\Phi}$.

For these reasons, we study the effect of \textit{varying} thermal boundary conditions by considering different configurations of the numerical domain -- in terms of both the thermophysical and geometric properties -- as presented in figure \ref{fig:schematic_configurations}. Commencing with the digital twin, we successively simplify the configuration. First, we replace the Newton cooling (NC) boundary condition \eqref{eq:BC_Newton_cooling} by another (stronger) Dirichlet condition \eqref{eq:BC_Dirichlet_top} (see panel (c)) similar to typical conjugate heat transfer (CHT) problems \citep{Perelman1961, Foroozani2021}. Second, we replace the glass top plate by an aluminium one such that $\kappa_{\textrm{st}} / \kappa_{\textrm{fl}} = \kappa_{\textrm{sb}} / \kappa_{\textrm{fl}}$ (see panel (d)). Third, the thickness of the top plate is adjusted to that of the bottom plate such that $\Gamma_{\textrm{st}} = \Gamma_{\textrm{sb}}$ (see panel (e)). Fourth and finally, we omit the solid top and bottom aluminium plates entirely and apply Dirichlet conditions (DIR) directly to the fluid layer (see panel (f)) -- this situation corresponds to $\left\{ \Gamma_{\textrm{st}}, \Gamma_{\textrm{sb}} \right\} \rightarrow 0$ and converged to or represents the traditional Rayleigh-Bénard convection configuration. 
Note that the different successive modifications build up on each other.

\newcommand{\hp}{\hphantom{1}}
\newcommand{\hpp}{\hphantom{.01}}
\begin{table}
\centering
\begin{tabular}{@{\hskip 0mm} l @{\hskip 2.0mm} c @{\hskip 2.0mm} c @{\hskip 2.0mm} r @{\hskip 2.0mm} l @{\hskip 2.0mm} l @{\hskip 2.0mm} l @{\hskip 2.0mm} l @{\hskip 2.0mm} l @{\hskip 0mm}}
\multicolumn{1}{l}{Code} &\multicolumn{1}{c}{$\Ra \quad$} & \multicolumn{1}{c}{$N_{\textrm{e}} \thickspace \thickspace$}	& \multicolumn{1}{c}{$\Bi \thickspace$}	& \multicolumn{1}{c}{$\Tinfty$}	& \multicolumn{1}{l}{$\quad \Tc$}	& \multicolumn{1}{c}{$\Tt$} & \multicolumn{1}{l}{$\quad \Tb$}	& \multicolumn{1}{l}{$\thickspace \thickspace \Th$} \\ [3pt]
\textcolor{gray}{2EXP}    & \textcolor{gray}{$2 \times 10^{5}$}		&                                        	&               & \textcolor{gray}{$-0.880$}     &                                                    &  \textcolor{gray}{$\scriptscriptstyle{\pm \mathcal{O} \left( 10^{-2} \right) }$}  &  \textcolor{gray}{$1 \thickspace \scriptscriptstyle{\pm \mathcal{O} \left( 10^{-2} \right) }$}  &  \textcolor{gray}{$1.2508$} \\
2NC5    & $2 \times 10^{5}$		& $85^{2} \times \left( 3 + 4 + 2 \right)$	&  $5.1$		& $-1.015$     & $-0.605 \hp \scriptscriptstyle{\pm \mathcal{O} \left( 10^{-4} \right) }$    &  $\scriptscriptstyle{\pm \mathcal{O} \left( 10^{-4} \right) }$  &  $1 \thickspace \scriptscriptstyle{\pm \mathcal{O} \left( 10^{-5} \right) }$  &  $1.0079$	\\
2NC6    & $2 \times 10^{5}$		& $85^{2} \times \left( 3 + 4 + 2 \right)$	&  $6.0$		& $-0.955$     & $-0.608 \hp \scriptscriptstyle{\pm \mathcal{O} \left( 10^{-4} \right) }$    &  $\scriptscriptstyle{\pm \mathcal{O} \left( 10^{-4} \right) }$  &  $1 \thickspace \scriptscriptstyle{\pm \mathcal{O} \left( 10^{-5} \right) }$  &  $1.0079$	\\
2CHTa   & $2 \times 10^{5}$		& $85^{2} \times \left( 3 + 4 + 2 \right)$	&               &              & $-0.603$                                           &  $\scriptscriptstyle{\pm \mathcal{O} \left( 10^{-4} \right) }$  &  $1 \thickspace \scriptscriptstyle{\pm \mathcal{O} \left( 10^{-5} \right) }$  &  $1.0079$	\\
2CHTb   & $2 \times 10^{5}$		& $85^{2} \times \left( 3 + 4 + 2 \right)$	&               &              & $-0.0035$                                          &  $\scriptscriptstyle{\pm \mathcal{O} \left( 10^{-5} \right) }$  &  $1 \thickspace \scriptscriptstyle{\pm \mathcal{O} \left( 10^{-5} \right) }$  &  $1.0079$	\\
2CHTc   & $2 \times 10^{5}$		& $85^{2} \times \left( 3 + 4 + 2 \right)$	&               &              & $-0.0079$                                          &  $\scriptscriptstyle{\pm \mathcal{O} \left( 10^{-5} \right) }$  &  $1 \thickspace \scriptscriptstyle{\pm \mathcal{O} \left( 10^{-5} \right) }$  &  $1.0079$	\\
2DIR    & $2 \times 10^{5}$		& $85^{2} \times \left( 3 + 4 + 2 \right)$	&               &              &                                                    &  $0$                                       &  $1$                                       &            \\
\\ 
\textcolor{gray}{4EXP}    & \textcolor{gray}{$4 \times 10^{5}$}		&                                        	&               & \textcolor{gray}{$-1.356$}     &                                                    &  \textcolor{gray}{$\scriptscriptstyle{\pm \mathcal{O} \left( 10^{-2} \right) }$}  &  \textcolor{gray}{$1 \thickspace \scriptscriptstyle{\pm \mathcal{O} \left( 10^{-3} \right) }$}  &  \textcolor{gray}{$1.2218$} \\
4NC6    & $4 \times 10^{5}$		& $125^{2} \times \left( 6 + 6 + 3 \right)$	&  $6.0$		& $-1.175$     & $-0.746 \thickspace \scriptscriptstyle{\pm \mathcal{O} \left( 10^{-4} \right) }$    &  $\scriptscriptstyle{\pm \mathcal{O} \left( 10^{-4} \right) }$  &  $1 \thickspace \scriptscriptstyle{\pm \mathcal{O} \left( 10^{-5} \right) }$  &  $1.0098$	\\
\\ 
\textcolor{gray}{7EXP}    & \textcolor{gray}{$7 \times 10^{5}$}		&                                        	&               & \textcolor{gray}{$-1.626$}     &                                                    &  \textcolor{gray}{$\scriptscriptstyle{\pm \mathcal{O} \left( 10^{-2} \right) }$}  &  \textcolor{gray}{$1 \thickspace \scriptscriptstyle{\pm \mathcal{O} \left( 10^{-3} \right) }$}  &  \textcolor{gray}{$1.2232$} \\
7NC6    & $7 \times 10^{5}$		& $125^{2} \times \left( 6 + 6 + 3 \right)$	&  $6.0$		& $-1.395$     & $-0.886 \thickspace \scriptscriptstyle{\pm \mathcal{O} \left( 10^{-4} \right) }$    &  $\scriptscriptstyle{\pm \mathcal{O} \left( 10^{-4} \right) }$  &  $1 \thickspace \scriptscriptstyle{\pm \mathcal{O} \left( 10^{-5} \right) }$  &  $1.0116$	\\
\end{tabular}
\caption{\justifying{Simulation parameters.
The Prandtl number $\Pr = 7.1$ and aspect ratio $\Gamma = 25$ with all walls of the closed domain obeying no-slip boundary conditions and lateral boundaries being perfectly insulated. 
The table contains beside the identifier further the Rayleigh number $\Ra$, the total number of spectral elements $N_{\textrm{e}} = N_{\textrm{e, x}} \times N_{\textrm{e, y}} \times \left( N_{\textrm{e, z, sb}} + N_{\textrm{e, z, fl}} + N_{\textrm{e, z, st}} \right)$ (with the polynomial order $N = 8$ on each spectral element, except for run 4NC6 where $N = 6$), the Biot number $\Bi$, as well as applied and resulting (spatio-temporally mean) temperatures at the different horizontal interfaces $\left\{ \Tinfty, \Tc, \Tt, \Tb, \Th \right\}$. 
Dynamically resulting temperatures are indicated by a quantification of the (temporal) standard deviation. 
The spatio-temporal average of $\Tt$ and $\Tb$ is typically $\mathcal{O} ( 10^{-4} )$ and $\mathcal{O} ( 10^{-5} )$ off its ideal value of $0$ and $1$, respectively.
The run time of all simulations $t_{\textrm{r}} = 12,000$ while the last $10,000$ have been used to gather results and statistical values. 
Motivating laboratory experiments are contrasted via rows with grey text colour while their printed temperatures assume ideal identifications of interface temperatures.
}}
\label{tab:simulation_parameters}
\end{table}
\let\hp\undefined
\let\hpp\undefined

\subsection{Comparison of different configurations of the numerical domain}
\label{subsec:Comparison_of_different_configurations_of_the_numerical_domain}
Table \ref{tab:simulation_parameters} summarises the final control parameters of each (production) simulation together with the spatial resolutions across the different parts of the domain. As some of the interface temperatures manifest dynamically, we include for those ones also the temporal standard deviation (around the spatio-temporal mean temperatures). 
After initialising the flows with these parameters, the long-living large-scale flow structures form and develop a statistically stationary pattern size within the first $2,000 \tauf$. This implies that also other global measures such as the heat and momentum transfer have converged \citep{Vieweg2021, Vieweg2022b, Vieweg2023a}.  We omit this transient period from our evaluation and run each simulation for additional $10,000 \tauf$ that will be analysed. Note that this runtime of the simulations exceeds the runtime of the laboratory experiments \citep{Moller2022a, Moller2022}.

\begin{figure}
\centering
\includegraphics[scale = 1.0]{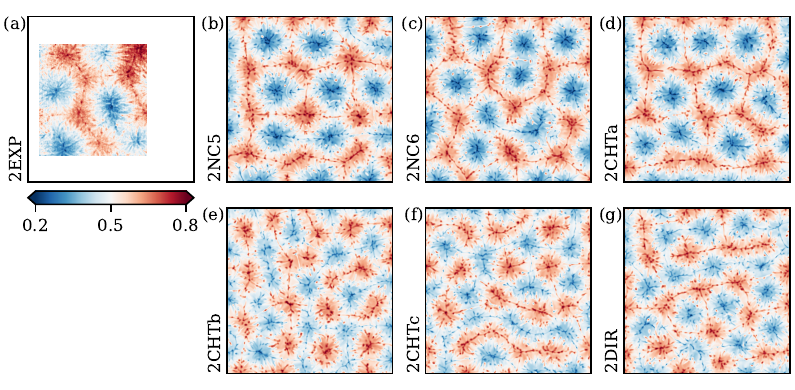}
\caption{\justifying{Flow structures at different thermal boundary conditions.
We visualise the instantaneous temperature field $T \left( x, y, z = 0.5, t = t_{\textrm{r}} \right)$ of (a) the laboratory experiment and (b -- g) each simulation at $\Ra = 2 \times 10^{5}$. The flow structures depend clearly on the thermal boundary conditions, see also figure \ref{fig:schematic_configurations}. The colour bar applies to all panels.}}
\label{fig:flow_structures_midplane_T}
\end{figure}

Figure \ref{fig:flow_structures_midplane_T} compares the different resulting flows by means of their final instantaneous temperature fields at mid-plane -- thermal boundary conditions clearly affect pattern formation. Structures become smaller for (stricter) conditions that are more similar to the plate-less Dirichlet configuration shown in panel (g), being in line with our previous studies \citep{Vieweg2021, Vieweg2022b, Vieweg2023a}. Vice versa, when this ideal configuration is successively left and the horizontal extent of the domain becomes smaller relative to the growing flow structures, the effect of the lateral side walls becomes stronger and they seem to impose preferential directions. This is most prominent in panels (b -- d) where the solid top plate is made of glass and so $\kappa_{\textrm{st}} / \kappa_{\textrm{fl}} \sim \mathcal{O} \left( 10^{0} \right)$. Note that it was not possible to measure these lateral wall-near regions in the motivating laboratory experiment \citep{Moller2022}, leading to the restricted field of view in panel (a). A comparison of this experimentally observed flow with the entire set of simulations confirms that the digital twins resemble the experiment best, particularly at $\Bi = 6.0$.

\newcommand{\hp}{\hphantom{1}}
\newcommand{\hpp}{\hphantom{.01}}
\begin{table}
\centering
\begin{tabular}{@{\hskip 0mm} l @{\hskip 3.2mm} l @{\hskip 3.2mm} l @{\hskip 3.2mm} c @{\hskip 3.2mm} c @{\hskip 3.2mm} c @{\hskip 3.2mm} c @{\hskip 0mm}}
\multicolumn{1}{l}{Code}	& \multicolumn{1}{r}{$\max \left( \Delta_{\textrm{hor}} \Tt \right) \quad$}	& \multicolumn{1}{l}{$\quad \thickspace \textrm{std} \left( \Tt \right)$}	& \multicolumn{1}{l}{$\quad \thickspace \Nu$} & \multicolumn{1}{l}{$\quad \Nuexp$}	& \multicolumn{1}{l}{$\quad \thickspace \thickspace \Re$}	& \multicolumn{1}{l}{$\quad \thickspace\Lambda_{T}$}	\\ [3pt]
\textcolor{gray}{2EXP}    & & & & \textcolor{gray}{$4.11 \pm 0.22$}	& \textcolor{gray}{$12.48 \pm 0.24$}	&  \textcolor{gray}{$7.53 \pm 0.22$}		\\
2NC5    & $0.598 \pm 0.016$     & $0.084 \pm \mathcal{O} \left( 10^{-4} \right)$	& $5.23 \pm 0.03$		& $5.27 \pm 0.05$	& $15.78 \pm 0.05$	&  $8.07 \pm 0.25$		\\
2NC6    & $0.586 \pm 0.017$     & $0.083 \pm \mathcal{O} \left( 10^{-4} \right)$	& $5.24 \pm 0.03$		& $5.33 \pm 0.05$	& $15.82 \pm 0.05$	&  $7.90 \pm 0.26$		\\
2CHTa   & $0.531 \pm 0.018$     & $0.073 \pm \mathcal{O} \left( 10^{-4} \right)$	& $5.22 \pm 0.03$		& $5.26 \pm 0.05$	& $15.75 \pm 0.06$	&  $7.59 \pm 0.21$		\\
2CHTb   & $0.007 \pm 0.000$     & $0.001 \pm \mathcal{O} \left( 10^{-5} \right)$	& $5.15 \pm 0.03$		& $5.19 \pm 0.05$	& $15.65 \pm 0.05$	&  $6.56 \pm 0.19$		\\
2CHTc   & $0.010 \pm 0.000$     & $0.002 \pm \mathcal{O} \left( 10^{-5} \right)$	& $5.15 \pm 0.03$		& $5.20 \pm 0.05$	& $15.66 \pm 0.05$	&  $6.62 \pm 0.19$		\\
2DIR    & $0$                   & $0$	                                            & $5.15 \pm 0.03$		& $5.19 \pm 0.05$	& $15.66 \pm 0.06$	&  $6.60 \pm 0.19$		\\
\\ 
\textcolor{gray}{4EXP}    & & & & \textcolor{gray}{$5.69 \pm 0.27$}	& \textcolor{gray}{$19.20 \pm 0.19$}	&  \textcolor{gray}{$7.69 \pm 0.27$}		\\
4NC6    & $0.561 \pm 0.017$     & $0.076 \pm \mathcal{O} \left( 10^{-4} \right)$	& $6.44 \pm 0.03$		& $6.49 \pm 0.07$	& $23.32 \pm 0.08$	&  $8.04 \pm 0.30$		\\
\\ 
\textcolor{gray}{7EXP}    & & & & \textcolor{gray}{$5.86 \pm 0.24$}	& \textcolor{gray}{$26.13 \pm 0.27$}	&  \textcolor{gray}{$7.77 \pm 0.53$}		\\
7NC6    & $0.541 \pm 0.018$     & $0.073 \pm \mathcal{O} \left( 10^{-4} \right)$	& $7.64 \pm 0.04$		& $7.68 \pm 0.09$	& $31.84 \pm 0.11$	&  $8.43 \pm 0.39$		\\
\end{tabular}
\caption{\justifying{Global characteristic measures 
of the simulations from table \ref{tab:simulation_parameters}. This table contains the maximum instantaneous temperature difference at the upper solid-fluid interface $\max \left( \Delta_{\textrm{hor}} \Tt \right)$, the instantaneous standard deviation of the temperature field at this interface $\textrm{std} \left( \Tt \right)$, the true global Nusselt number $\Nu$ (which includes the diffusive heat transport), the experimentally accessible Nusselt number $\Nuexp$, the Reynolds number $\Re$, as well as the integral length scale of the temperature field $\Lambda_{T}$. All values are provided as temporal means together with the corresponding standard deviation.
Motivating laboratory experiments are contrasted via rows with grey text colour and are based on a restricted field.
Revised values of their $\Nuexp$ and $\Re$ are reported in section \ref{subsec:Reassessment_the_original_laboratory_measurement_data}.
}}
\label{tab:simulation_outcome}
\end{table}
\let\hp\undefined
\let\hpp\undefined

Extending this first visual or qualitative impression, we proceed by quantifying various measures of the flows. 
Foremost, we consider the largest instantaneous horizontal temperature difference 
\begin{equation}
\label{eq:def_max_Delta_hor_Tt}
\max \left( \Delta_{\textrm{hor}} \Tt \right) \thickspace (t)
= \max_{x, y} \left( \Tt \right) - \min_{x, y} \left( \Tt \right) 
\end{equation}
at the (proner) upper solid-fluid interface. 
Secondly, this is complemented by the instantaneous standard deviation $\textrm{std} \left( \Tt \right)$. These two quantities are used to probe the inhomogeneity of the temperature field at this interface.
Thirdly, we quantify quantify the ratio of the (average) total heat current $\bm{J} = \bm{u} T + \bm{J}_{\textrm{dif}}$ across the fluid layer to the diffusive heat current $\bm{J}_{\textrm{dif}} = - \nabla T / \sqrt{\Ra \Pr}$ that took place in the case of pure heat conduction by the (global) Nusselt number \citep{Otero2002, Vieweg2023a}
\begin{equation}
\label{eq:def_Nusselt_number}
\Nu \thickspace (t)
= \frac{\langle \bm{J} \cdot \bm{e}_{z} \rangle_{V}}{\langle \bm{J}_{\textrm{dif}} \cdot \bm{e}_{z} \rangle_{V}} 
= \langle - \frac{\partial T }{\partial z} + \sqrt{\Ra \Pr} \thickspace u_{z} T \rangle_{V}
= 1 + \sqrt{\Ra \Pr} \thickspace \langle u_{z} T \rangle_{V} .
\end{equation}
Fourthly, we assess the momentum transport using the Reynolds number \citep{Scheel2017}
\begin{equation}
\label{eq:def_Reynolds_number}
\Re \thickspace (t)
:= \sqrt{\frac{\Ra}{\Pr}} \thickspace u_{\textrm{rms}} \qquad \textrm{with } u_{\textrm{rms}} := \sqrt{ \langle \bm{u}^{2} \rangle_{V} } .
\end{equation}
Finally, the so-called integral length scale \citep{Parodi2004} 
\begin{equation}
\label{eq:def_integral_length_scale}
\Lambda_{T} \left( z = 0.5, t \right) 
:= 2 \pi \frac{\int_{k_{\textrm{h}}} \left[ E_{TT} / k_{\textrm{h}} \right] dk_{\textrm{h}}}{\int_{k_{\textrm{h}}} E_{TT} \thickspace dk_{\textrm{h}}}
\end{equation}
is used to measure the present characteristic pattern size. $E_{TT} \equiv E_{TT} \left( k_{\textrm{h}}, z = 0.5, t \right)$ represents the azimuthally averaged Fourier energy spectrum of the temperature field and $k_{\textrm{h}}$ the horizontal wave number \citep{Vieweg2023a}. 
All of these quantities are summarised and contrasted with respect to their temporal mean value and standard deviation in table \ref{tab:simulation_outcome}. 

Our quantification of thermal inhomogeneities highlights prominently that the upper solid-fluid interface can become strongly inhomogeneous depending on the choice of the plate material. Given a glass plate with $\kappa_{\textrm{st}} / \kappa_{\textrm{fl}} \sim \mathcal{O} \left( 10^{0} \right)$, the horizontal temperature difference can reach $60 \%$ of the temperature drop across the Rayleigh-Bénard convection layer. In contrast, this is reduced to $1 \%$ once the plate is made of aluminium with $\kappa_{\textrm{st}} / \kappa_{\textrm{fl}} \sim \mathcal{O} \left( 10^{2} \right)$. This observation is qualitatively confirmed by the corresponding standard deviation of the interface temperature and can directly be related to the different ratios of thermal diffusion time scales $\tau_{\kappa \textrm{, fl}} / \tau_{\kappa \textrm{, st}} \equiv \kappa_{\textrm{st}} / \kappa_{\textrm{fl}}$. 

\begin{figure}
\centering
\includegraphics[scale = 1.0]{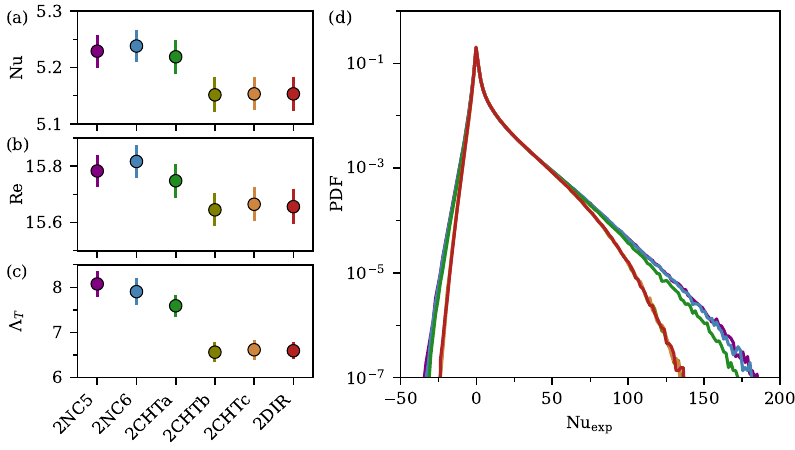}
\caption{\justifying{Effect of different (partly non-ideal) thermal boundary conditions.
While the global (a, b) heat and momentum transport depend only weakly on the configuration at $\Ra = 2 \times 10^{5}$, (c) the size of the large-scale flow structures is strongly influenced. Error bars depict the standard deviation, see table \ref{tab:simulation_outcome}.
In contrast to its global measure, (d) the statistical distribution of the local heat transport depends sensitively on the thermal boundary conditions.
}}
\label{fig:global_characteristic_measures_and_PDF_of_Nuexp}
\end{figure}

Figure \ref{fig:global_characteristic_measures_and_PDF_of_Nuexp} visualises the trends in the other three quantities with the applied thermal boundary conditions in panels (a -- c). 
We find that both the global heat and momentum transport are enhanced by approximately $1 \%$ when the configuration comprises a glass top plate. The origin of this intensification can be found in the enhanced buoyancy $\bm{b} = T \bm{e}_{z}$, the latter of which becomes possible due to the looser bound on the temperature field (here with $T < 0$ possible) when thermal inhomogeneities become significant at the boundaries.
However, their impact on $\Nu$ and $\Re$ is tiny compared to an average $19 \%$ increase in the size of long-living large-scale flow structures as measured by $\Lambda_{T}$. 
This underlines the effect of thermal boundary conditions or inhomogeneities at the boundaries on pattern formation. 

Unfortunately, the experimentally present measurement techniques did not admit to determine $\partial T  / \partial z$ as part of $\Nu$. We thus extend our analysis of the (local) heat transfer to 
\begin{equation}
\label{eq:def_Nusselt_number_experiment}
\Nuexp \thickspace ( x, y, z = 0.5, t)
:= \sqrt{\Ra \Pr} \thickspace u_{z} \Theta \qquad \textrm{with } \Theta := T - T_{\textrm{lin}} 
\end{equation}
where $\Theta$ represents the temperature deviation field around the linear conduction profile $T_{\textrm{lin}} := \langle \Tb - \left( \Tb - \Tt \right) z \rangle_{A} \simeq 1 - z$. This experimentally accessible Nusselt number is one subset of the (true) Nusselt number introduced in equation \eqref{eq:def_Nusselt_number}, and its mean value can (provided $\langle \partial T / \partial z \left( z = 0.5 \right) \rangle_{A} = 0$ and certain further criteria) coincide with that of $\Nu$ as described in more detail in \citep{Vieweg2023}. 

We contrast the time-averaged probability density functions (PDFs) of $\Nuexp$ in figure \ref{fig:global_characteristic_measures_and_PDF_of_Nuexp} (d) for the entire set of considered domain configurations. Note that $\Nuexp > 0$ corresponds to regions of \textit{intended} heat transfer with either $u_{z} > 0$ and $\Theta > 0$ or $u_{z} < 0$ and $\Theta < 0$, whereas $\Nuexp < 0$ corresponds to regions of \textit{inverted} heat transfer with either $u_{z} > 0$ and $\Theta < 0$ or $u_{z} < 0$ and $\Theta > 0$.
In contrast to the global values of $\Nu$ (see again panel (a)), we find that the statistical distribution of the (local) heat transfer depends sensitively on the thermal boundary conditions. The tails, especially the positive ones which comprise the very cold down-welling fluid, are significantly enhanced when thermal inhomogeneities are allowed at the boundaries. The PDF becomes thus wider as bounds on $T$ get looser.

We conclude this section by comparing our numerical configurations with data from the experiment, the latter of which we therefore include in tables \ref{tab:simulation_parameters} and \ref{tab:simulation_outcome}. 
We find that both $\Nuexp$ and $\Re$ are smaller or seem to be underestimated in the laboratory experiment. In contrast, $\Lambda_{T}$ agrees well once a glass top plate is considered in the simulations. In other words, the numerical inclusion of solid plates with realistic thermophysical and geometrical properties allows to explain the increased size of the flow pattern in the experiment. As the digital twin agrees best with the experiment when $\Bi = 6.0$, we will focus on this parameter in the following.

\section{The impact of the experimental measurement procedure}
\label{sec:impact_of_experimental_measurement_procedure}

As our digital twin with $\Bi = 6.0$ successfully explains the increased flow pattern size from the experiment at $\Ra = 2 \times 10^{5}$, we extend it towards the larger experimentally provided $\Ra = \left\{ 4, 7 \right\} \times 10^{5}$ \citep{Moller2022a, Moller2022} and include the corresponding data in tables \ref{tab:simulation_parameters} and \ref{tab:simulation_outcome}.

A comparison of the resulting size of large-scale flow structures confirms a good agreement between the numerical and experimental flows across the entire range of Rayleigh numbers, especially when keeping the standard deviation and limited field of view for the experimental data in mind. However, we find that the overall heat and momentum transfer persist to disagree strongly between both approaches. Both $\Nuexp$ and $\Re$ seem to be underestimated by roughly $20 \%$ in the experiment. 
Hence, we proceed by implementing and analysing the detailed effects of the experimentally present measurement procedure. 

\subsection{Numerical implementation of measurement errors and uncertainties}
\label{subsec:Numerical_implementation_of_measurement_errors_and_uncertainties}

We emulate the effects of experiment measurements by considering the following aspects:
\begin{enumerate}[nosep]
    \item The neglect of the vertical temperature gradient (i.e., affecting $\Nu$ or $\Nuexp$).
    \item Any spatial averaging (affecting $u_{z}$ and $T$), both
    \begin{enumerate}[nosep]
        \item vertically across a slab due to the light sheet thickness and
        \item horizontally in interrogation windows as required for particle image velocimetry and thermometry (PIV and PIT, respectively).
    \end{enumerate}
    \item Thermal measurement deviations (only affecting $T$) associated with 
    \begin{enumerate}[nosep]
        \item an erroneous determination of plate temperatures from only a few point measurements and
        \item uncertainties from the colour identification of thermochromic liquid crystals (TLCs).
    \end{enumerate}
\end{enumerate}

For the flows at hand, the systematic omissions of (1) and (2.a) affect $\Nu$ by only about $1 \%$ and are thus considered to be negligible. As this agrees with \citep{Vieweg2023}, we disregard these two effects and focus on the effect of the remaining three ones on $\Nuexp$ in the following.

We will apply a horizontal averaging in interrogation windows of an approximate size of (depending on $\Ra$) $0.11^{2}$ or $0.17^{2}$ \citep{Moller2022a, Moller2022} in dimensionless spatial coordinates to both $\bm{u}$ and $T$ -- this blurs those fields essentially. Furthermore, we manipulate the temperature field to incorporate the outlined systematic errors as well as random uncertainties. This is realised as follows.

Let 
\begin{equation}
\label{eq:measured_nondimensional_T_origin}
\tilde{T}^{\textrm{m}} = \frac{T^{\textrm{m}} - \langle \Tt^{\textrm{m}} \rangle_{A, t}}{\Delta T^{\textrm{m}}} \qquad \textrm{with } \Delta T^{\textrm{m}} = \langle \Tb^{\textrm{m}} - \Tt^{\textrm{m}} \rangle_{A, t}
\end{equation}
be the resulting or perceived non-dimensional temperature for some \textit{measured} (i.e., dimensional) local temperature value $T^{\textrm{m}}$. Any of the involved measured temperatures $\left\{ T^{\textrm{m}}, \Tb^{\textrm{m}}, \Tt^{\textrm{m}} \right\}$ can be subject to individual errors and uncertainties via 
\begin{equation}
\label{eq:def_general_measurement_error}
T_{\Phi}^{\textrm{m}} = T_{\Phi} + \delta T_{\Phi}
\end{equation}
with $T_{\Phi}$ representing the true value and $\delta T_{\Phi}$ the measurement deviation. This allows to conclude that the perceived non-dimensional temperature is related via 
\begin{equation}
\label{eq:measured_nondimensional_T_based_on_nondimensional_errors}
\tilde{T}^{\textrm{m}} 
= \frac{\tilde{T} + \delta \tilde{T} - \langle \delta \tilde{\Tt} \rangle_{A, t}}{\langle 1 + \delta \tilde{\Tb} - \delta \tilde{\Tt} \rangle_{A, t}} 
\equiv \underbrace{\frac{\tilde{T} - \langle \delta \tilde{\Tt} \rangle_{A, t}}{\Delta \tilde{T}^{\textrm{m}}}}_{\mathclap{\begin{subarray}{c} \textrm{error solely due to} \\ \textrm{plate temperatures} \end{subarray}}} 
\thickspace + \thickspace 
\underbrace{\frac{\delta \tilde{T}}{\Delta \tilde{T}^{\textrm{m}}}}_{\mathclap{\begin{subarray}{c} \textrm{uncertainty} \\ \textrm{due to TLCs} \end{subarray}}}
\qquad \textrm{with } \Delta \tilde{T}^{\textrm{m}} = \langle 1 + \delta \tilde{\Tb} - \delta \tilde{\Tt} \rangle_{A, t}
\end{equation}
to the corresponding non-dimensional measurement deviation $\delta \tilde{T}_{\Phi} := \delta T_{\Phi} / \Delta T$, the latter of which are defined based on the true temperature difference across the fluid layer $\Delta T$. In other words, we derived a framework to add non-dimensional measurement deviations $\delta \tilde{T}_{\Phi}$ to true non-dimensional values $\tilde{T}$. 
Note that $\delta \tilde{T}_{\Phi} > 0$ implies that the perceived value is larger than the true value, see equation \eqref{eq:def_general_measurement_error}. 

Our approach allows to disentangle the thermal measurement deviations $\left\{ \delta \tilde{T}, \delta \tilde{\Tb}, \delta \tilde{\Tt} \right\}$ depending on their origin as shown on the right of equation \eqref{eq:measured_nondimensional_T_based_on_nondimensional_errors}. 
The amplitude of the TLC-related $\delta T$ has been quantified in \citet{Moller2022} as a function of the true temperature, i.e., we are given $\delta T = \delta T \left( T \right)$. We can make use of this relation in equation \eqref{eq:measured_nondimensional_T_based_on_nondimensional_errors} via $\delta T / \Delta T^{\textrm{m}} \equiv \delta \tilde{T} / \Delta \tilde{T}^{\textrm{m}}$ -- the associated standard deviation is on average $6 \%$ of the perceived temperature drop across the fluid layer, but (depending on $T$ and $\Ra$) the local value can easily exceed $10 \%$. We model this random local uncertainty numerically as Gaussian noise.
In contrast, the experimental data does not allow for an estimation of any systematic errors associated with the determination of the plate temperatures, $\langle \delta \tilde{\Tb} \rangle_{A, t}$ and $\langle \delta \tilde{\Tt} \rangle_{A, t}$.

\begin{figure}
\centering
\includegraphics[scale = 1.0]{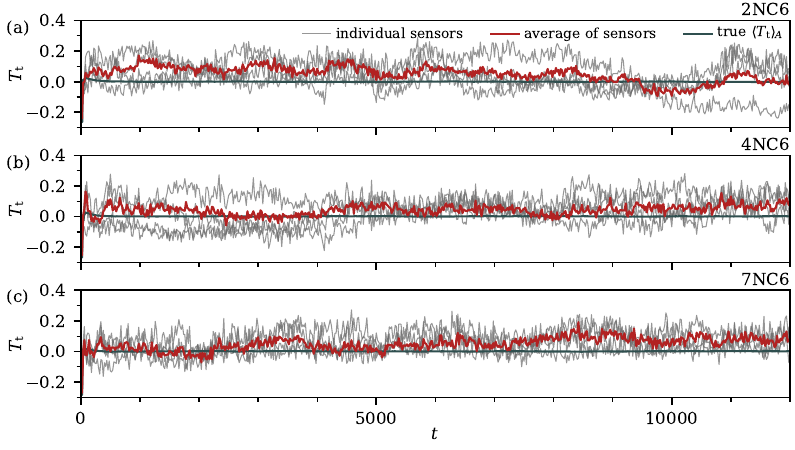}
\caption{\justifying{Numerical top (glass) plate temperature measurement.
Although the true mean interface temperature $\langle \Tt \rangle_{A} = 0$ already after $\mathcal{O} \left( 10^{2} \tauf \right)$, four point sensors are too few to identify it accurately.
}}
\label{fig:evolution_plate_measurement}
\end{figure}

We therefore estimate the latter based on a numerical imitation of the experimental plate temperature measurement. In the experiment \citep{Moller2022}, $\langle \Tb \rangle_{A}$ is determined based on $5$ temperature probes within the bottom solid plate, whereas $\langle \Tt \rangle_{A}$ is determined based on $4$ sensors that are glued onto the top plate. Resembling this process with our digital twin, we find that this technique allows capturing the bottom plate's mean temperature almost perfectly due to its homogeneous temperature distribution -- the error is of order $\mathcal{O} \left( 10^{-4} \right)$. 
This changes once the top plate with its thermal inhomogeneities is considered. Figure \ref{fig:evolution_plate_measurement} tracks therefore the evolution of the temperature signals at the upper solid-fluid interface. Although $\langle \Tt \rangle_{A} = 0$ already shortly after the initialisation, the local temperature signals fluctuate strongly. An arithmetic average dampens these fluctuations only to a certain extent. Crucially, even a time-average of the instantaneous ensemble-average does not yield the correct mean interface temperature -- instead, we find deviations of approximately $5 \%$ which is roughly similar to $\langle \textrm{std} \left( \Tt \right) \rangle_{A, t}$. 
Moreover, we find that the standard deviations agree with that of the experimentally obtained time series \citep{Moller2022}. 
This analysis highlights that the temperature at the thermally inhomogeneous top plate varies strongly over space and time, and so $4$ sensors are too few to identify the mean temperature at the solid-fluid interface accurately. The derived non-dimensional temperatures in the experiment might thus be biased. 
In the following, we will drop the tildes and assume $\langle \delta \Tb \rangle_{A, t} = 0$ and $\langle \delta \Tt \rangle_{A, t} = \langle \textrm{std} \left( \Tt \right) \rangle_{A, t}$.

\subsection{The Impact of measurement deviations on the (local) heat transfer}
\label{subsec:The_impact_of_measurement_deviations_on_the_local_heat_transfer}
So far, we have described the origin of different experimentally present measurement errors and uncertainties, quantified their individual size, and derived a framework to correspondingly modify numerical data. We proceed by adding key measurement effects to the numerical data at $\Ra = 2 \times 10^{5}$ and analysing their effect on both the statistical and mean heat transfer in more detail.

\begin{figure}
\centering
\includegraphics[scale = 1.0]{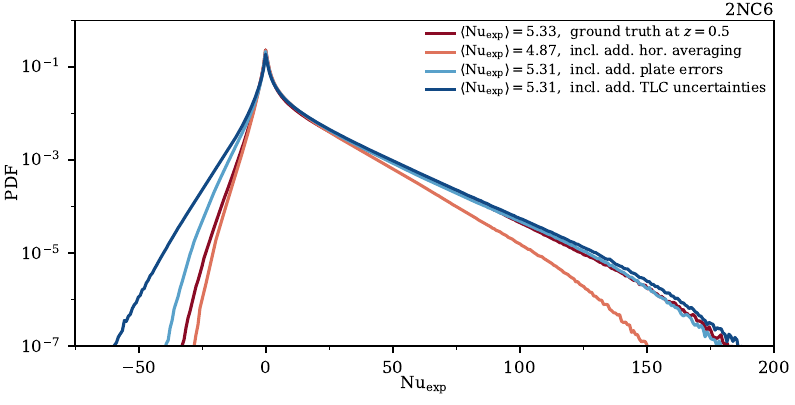}
\caption{\justifying{Impact of the measurement procedure.
Although the latter affects the perceived statistical distribution of the local heat transfer significantly, its mean value $\langle \Nuexp \rangle$ seems almost unchanged.
Note that the different contributions outlined in the legend are applied cumulatively.
}}
\label{fig:measurement_uncertainties_PDF_of_Nuexp}
\end{figure}

Figure \ref{fig:measurement_uncertainties_PDF_of_Nuexp} visualises this iterative process, the latter of which starts with the ground truth of $\Nuexp$ as defined in equation \eqref{eq:def_Nusselt_number_experiment}. Note that this ground truth is based on the unmodified numerical fields and already known from figure \ref{fig:global_characteristic_measures_and_PDF_of_Nuexp} (d). 
In analogy to the PIV and PIT processing, we start by incorporating a horizontal averaging in interrogation windows. Since any spatial averaging dampens local extrema, the PDF's tails become weaker and it narrows significantly. Importantly, also the associated average decreases by $9 \%$.
Next, we start perceiving the mean temperature of the upper solid-fluid interface hotter than it actually is. This leads to cold temperatures appearing even colder and so the range of observed temperature values broadens. As a consequence, the PDF tails become stronger and the associated average increases by about $9 \%$.
Including eventually also the TLC-related uncertainties spreads the tails of the PDF beyond any of the previously plotted ones. This affects the weaker negative tails more strongly than the stronger positive ones due to the intricate composition of $\Nuexp$, see again section \ref{subsec:Comparison_of_different_configurations_of_the_numerical_domain}. Interestingly, the mean perceived heat transfer across the fluid layer is not affected despite the complex relation between $\delta T$ and $T$.

After considering all these different aspects, $\langle \Nuexp \rangle$ is reduced by less than $1 \%$ compared to its original value. However, the statistical heat transfer has been affected strongly in a way that depends sensitively on all of them.

\subsection{Comparison of laboratory measurements with manipulated numerical data}
\label{subsec:Comparison_of_laboratory_measurements_with_manipulated_numerical_data}
The previous sections \ref{subsec:Comparison_of_different_configurations_of_the_numerical_domain} and \ref{subsec:The_impact_of_measurement_deviations_on_the_local_heat_transfer} have laid out the foundations to understand the detailed effects of experimentally present non-ideal thermal boundary conditions and uncertainties introduced by the measurement techniques, respectively. 
In this section, we will consider \textit{perceived}, i.e., manipulated numerical quantities only and so we omit any related superscript. 

\begin{figure}
\centering
\includegraphics[scale = 1.0]{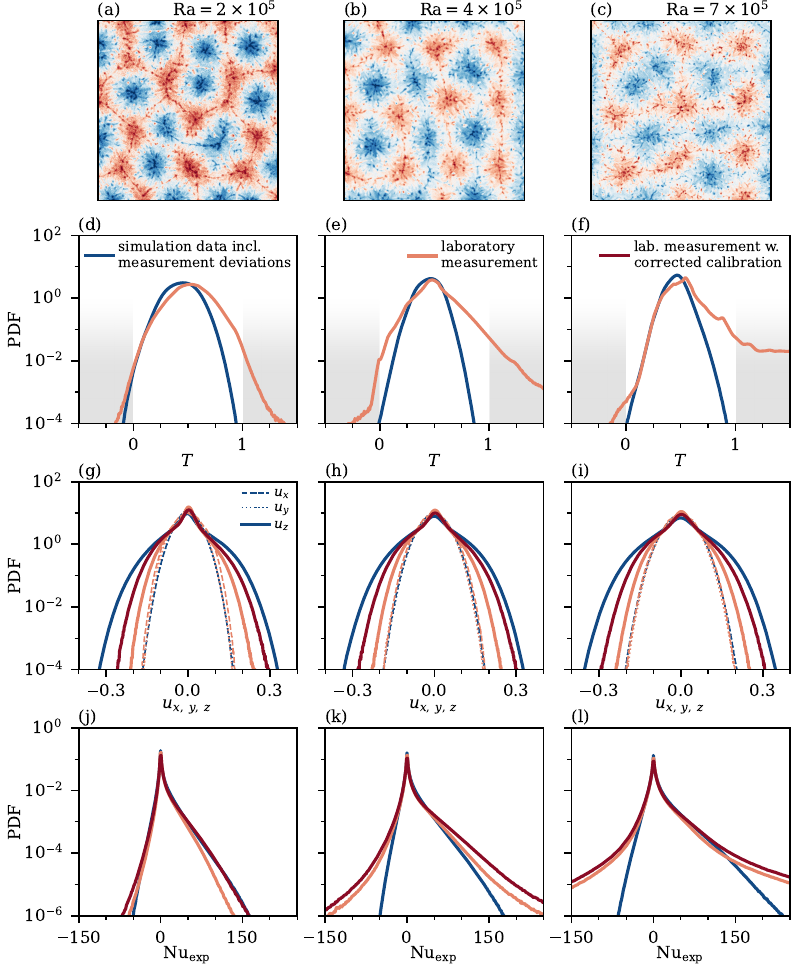}
\caption{\justifying{Contrast of statistical data obtained from simulations and experiments.
Here we exploit simulation data at $\Bi = 6.0$ only -- see panels (a -- c) for instantaneous temperature fields $T \left( x, y, z = 0.5, t = t_{\textrm{r}} \right)$ which are subjected to experiment-like measurement deviations. The colour map coincides with figure \ref{fig:flow_structures_midplane_T} and the corresponding $\Ra$ are given at the top.
Below, the statistical distribution of the (d -- f) temperature $T$, (g -- i) velocities $u_{x, y, z}$, and (j -- l) experimentally accessible Nusselt number $\Nuexp$ are contrasted with laboratory experimental data. 
Although the $\textrm{PDFs} \left( T, u_{x, y} \right)$ seem to agree well between simulations and experiments, $u_{z}$ appears to be underestimated in the case of the latter across all $\Ra$.
The correction of the subsequently discovered calibration mistake allows for an improved convergence of results.
The second row defines the colour encoding for all the PDFs, whereas its grey backgrounds indicate measurements that used to be discarded in past experiments.
}}
\label{fig:contrast_of_statistical_data_from_simulations_and_experiment}
\end{figure}

Figure \ref{fig:contrast_of_statistical_data_from_simulations_and_experiment} (a) presents the resulting instantaneous temperature field $T \left( x, y, z = 0.5, t = t_{\textrm{r}} \right)$ from the numerical simulation at $\Ra = 2 \times 10^{5}$ which mimics the experimental measurement procedure. A comparison with figure \ref{fig:flow_structures_midplane_T} (c) highlights the impact of the latter. 
The perceived temperature fields at larger Rayleigh numbers are included in figures \ref{fig:contrast_of_statistical_data_from_simulations_and_experiment} (b, c). 
Although the enhanced turbulence results in an increased mixing of the scalar temperature, the decreasing ratio of the pattern size to the horizontal extent of the domain, i.e., $\Lambda_{T} / \Gamma$, leads to a stronger influence of side walls on pattern formation -- the long-living large-scale flow structures thus tend to align with the sidewalls.

Figure \ref{fig:contrast_of_statistical_data_from_simulations_and_experiment} (d -- f) contrasts the perceived temperature distributions from the numerical simulations and laboratory experiments by means of their time-averaged PDFs. 
Note that we observe temperatures beyond the range $\left[ 0, 1 \right]$ due to two independent reasons. Firstly, the non-ideal thermal boundary conditions allow for spatial variations of the temperature field at the fluid-solid interfaces. With $\kappa_{\textrm{sb}} / \kappa_{\textrm{fl}} \sim \mathcal{O} \left( 10^{2} \right) \gg \kappa_{\textrm{st}} / \kappa_{\textrm{fl}} \sim \mathcal{O} \left( 10^{0} \right)$, this affects practically solely the lower limit of $T$ (see again section \ref{subsec:Comparison_of_different_configurations_of_the_numerical_domain}). Secondly, the uncertainties associated with the TLC measurements cause the detection of temperatures beyond both sides of $\left[ 0, 1 \right]$. Any ramification on either of these bounds depends on the functional dependence $\delta T = \delta T \left( T \right)$. 
As a result, we detect in both the modified numerical as well as the experimental data temperatures beyond the spatio-temporal averages $\langle \Tt \rangle_{A, t} = 0$ and $\langle \Tb \rangle_{A, t} = 1$.

The perceived numerical data shows slightly stronger tails for lower temperatures compared to the higher ones. We find that these left tails, as well as the peaks, coincide very well with the experimental data, see in particular panel (f). However, the laboratory measurements show an over-representation of larger temperatures, especially at larger $\Ra$. 
This is related to locally increased uncertainties of the TLCs at higher temperatures and observation angles \citep{Konig2019, Moller2019, Moller2021, Moller2022}. For this reason, previous studies have disregarded any temperature measurements outside the range $\left[ 0, 1 \right]$ (indicated by the grey shaded areas in figures \ref{fig:contrast_of_statistical_data_from_simulations_and_experiment} (d -- f) ) \citep{Moller2022}. 
If the situation was symmetric, this could be considered reasonable. 
However, as this ignores the natural asymmetry due to the thermal boundary conditions and TLC uncertainties, and to avoid sharp cut-offs at or due to large temperatures, we will retain those measurements in this study.

Figure \ref{fig:contrast_of_statistical_data_from_simulations_and_experiment} (g -- i) opposes the perceived velocity components with respect to their time-averaged PDFs. 
An almost perfect match of the horizontal velocities $u_{x, y}$ between the numerical and experimental approach confirms the good resemblance of the latter by its digital twin.
The buoyancy-driven convective heat transport induces vertical velocities that exceed the horizontal ones. Although this is true for both data sources, vertical velocities from the simulations are significantly stronger compared to those from the experiments. Since this cannot be resolved by our digital twin and the applied modifications, it suggests that the vertical or out-of-plane component of the velocity might have been systematically (i.e., independently of $\Ra$) underestimated by the stereoscopic PIV measurements.

Building up on the above insights, figure \ref{fig:contrast_of_statistical_data_from_simulations_and_experiment} (j -- l) compares the resulting experimentally accessible Nusselt number $\Nuexp$.
On the one hand, we find that the PDFs offer a similar shape in panel (j) with a growing discrepancy towards larger $\Ra$. This latter circumstance suggests that the over-representation of those tails can be attributed to the $\Ra$-dependent increase of uncertainties associated with the TLC measurements, see again panels (d -- f). 
On the other hand, we find $\langle \Nuexp \rangle_{A, t} = \left\{ 5.31, 6.18, 7.56 \right\}$ based on the manipulated numerical data. Since these values are independently of $\Ra$ larger than the experimentally observed ones (see again table \ref{tab:simulation_outcome}), this supports the suspicion of an experimental underestimation of the vertical velocity (see again panels (g -- i)).

\subsection{Re-assessment of the original laboratory measurement data}
\label{subsec:Reassessment_the_original_laboratory_measurement_data}
Even after resembling the laboratory measurement procedure, our digital twins offer an increased global heat transfer compared to the experiment. As our data suggests an experimental underestimation of $u_{z}$, $\langle \Nuexp \rangle_{A, t} \sim u_{z}$ appears to be roughly $22 \%$ larger in the case of the former compared to the latter. 

We therefore carefully scrutinise or re-assess the processing of the original stereo-PIV measurement data starting from the raw uncalibrated camera images. This reveals that the previously used pinhole camera calibration model with a subsequent self-calibration \citep{Wieneke2005} did not account for the optical refraction effect between the cameras and the fluid layer, resulting in wrong vertical distances of the calibration planes and thus a systematic underestimation of only the vertical or out-of-plane velocity component. 
This issue can be eliminated by using a polynomial calibration which creates the image-to-world mapping by fitting polynomials through the calibration markers. This approach does not require any modeling of the optical path and is well-suited for complex set-ups. As a downside, it allows for limited extrapolations beyond the calibrated region only which is not of relevance for us anyway.
We thus re-process the PIV data using a polynomial calibration and find a relative increase of $u_{z}$ by approximately $24 \%$ for all different Rayleigh numbers. 
As this number propagates directly to $\Nuexp$, they agree now almost perfectly with the expectations based on the digital twin. This re-assessment reveals new (corrected) global measures of heat and momentum transport in the laboratory experiment of $\Nuexp = \left\{ 5.09, 7.04, 7.25 \right\}$ and $\Re = \left\{  13.56, 20.99, 28.55 \right\}$, respectively. 
Figure \ref{fig:contrast_of_statistical_data_from_simulations_and_experiment} (g -- l) includes also the corrected statistical data. 

\begin{figure}
\centering
\includegraphics[scale = 1.0]{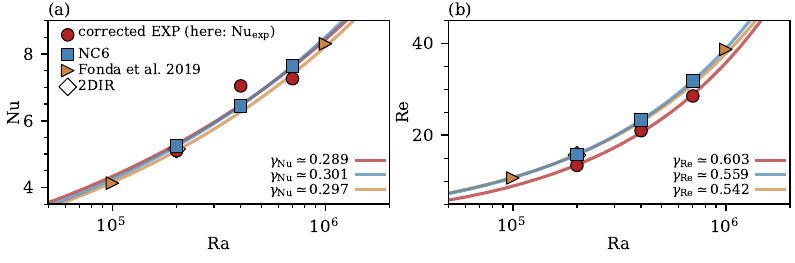}
\caption{\justifying{Scaling of the global heat and momentum transfer.
We contrast available experimental and numerical data -- that offer both $\Pr \simeq 7$ and a closed Cartesian domain with $\Gamma = 25$ -- between (i) idealised constant temperature and (ii) experiment-like conditions. 
Markers specify the exact data points whereas solid lines represent the resulting (extrapolated) fitted curves with the corresponding scaling exponents provided in the legends.
\citep{Fonda2019} offers numerical data at constant temperature conditions.
}}
\label{fig:scaling_of_Nu_and_Re}
\end{figure}

Finally, it is certainly of interest how these corrected values and the thermal boundary conditions in more general affect the overall scaling of the global heat and momentum transfer. We thus conclude with a detailed comparison of $\Nu$ and $\Re$ in figure \ref{fig:scaling_of_Nu_and_Re}. 
First, we find that the corrected data points from the experiment offer an improved conformity with its digital twin NC6. While $\Nuexp$ at $\Ra = 4 \times 10^{5}$ seems to stand out in this direct comparison, we can trace this back to the usage of a different temperature calibration for this single experiment run. 
Nevertheless, we find that the resulting scaling exponents describing $\Nu \sim \Ra^{\gamma_{\Nu}}$ \citep{Plumley2019, Vieweg2023a} and $\Re \sim \Ra^{\gamma_{\Re}}$ are quite similar.
This confirms that the underlying physics of the flow is properly captured by the experimental measurement data and that this physics's detection is mostly unaffected by the measurement deviations.
Second, contrasting the digital twin NC6 with numerical data at constant temperature boundary conditions from \citep{Fonda2019} and our simulation 2DIR underlines the marginal effect of variations of thermal boundary conditions (as far as considered in this study) on both quantities. We find the scaling exponents to coincide virtually and the resulting fitted curves to be almost congruent. This comparison confirms our results from section \ref{subsec:Comparison_of_different_configurations_of_the_numerical_domain} at $\Ra = 2 \times 10^{5}$ and extends them across the range $\Ra = \left[ 2, 7 \right] \times 10^{5}$. 

\section{Discussion and perspective}
\label{sec:Discussion_and_perspective}
This comparative study has systematically scrutinised discrepancies between observations from numerical and experimental approaches. In particular, the long-living large-scale flow structures \citep{Vieweg2023, Vieweg2023a, Vieweg2024a} seemed to show an increased characteristic size but decreased induced heat transfer in an experimental approach relative to numerical ones. As constraints emerging from the measurement techniques limit corresponding modifications to the experiment set-up, we decided to shift perspectives and use simulations instead.

Both the horizontal extent of the domain as well as the interaction of the fluid layer with the adjacent solid plates are of crucial significance for the formation of flow structures and thus their induced heat transfer \citep{Krug2020, Vieweg2021}. Past numerical studies have not accounted for these two aspects simultaneously \citep{Czarnota2013, Foroozani2021, Pandey2018, Vieweg2021, Vieweg2023}.
Hence, we created a digital twin of the laboratory experiment by including the solid plates with respect to their geometry, thermophysical properties, and respective external thermal boundary conditions.

We find for this twin that $\max \left( \Delta_{\textrm{hor}} \Tt \right) > 0.5$ and thus exceeds half of the total temperature drop $\dT$ across the fluid layer for all $\Ra$. This shows that the upper solid-fluid interface is extremely prone to thermal inhomogeneities and proves that the past assumption \enquote{the isothermal boundary conditions can be considered as fulfilled in good approximation} \citep{Moller2020} has to be reconsidered for the experimentally covered range of $\Ra$. 
Only the bottom aluminium plate renders the corresponding interface iso-thermal.
A comparison of the different convection and heat transfer coefficients $\left\{ h_{\Bi}, h_{\textrm{st}}, h_{\textrm{fl}}, h_{\textrm{sb}} \right\}$ across the different layers (see again equations \eqref{eq:IC_NewtonBC_D_hBi} and \eqref{eq:IC_NewtonBC_h}) shows that thermal conduction across the top plate represents the essential bottleneck for heat transfer in the domain and thus alters the boundary conditions significantly. For instance, $h_{\textrm{st}} / h_{\textrm{fl}} \approx 1.6$ in contrast to $h_{\Bi} / h_{\textrm{fl}} \approx 2.9$ and $h_{\textrm{sb}} / h_{\textrm{fl}} \approx 126$ for $\Ra = 2 \times 10^{5}$. 
As is underlined by simulation 2CHTa (corresponding to $h_{\Bi} \rightarrow \infty$), the effect of increasing $\Bi$ is very limited. 

A systematic step-wise simplification of the digital twin to the standard numerical set-up shows that realistic thermophysical properties do explain the experimentally observed increased structure size $\Lambda_{T}$ -- confirming our previous results \citep{Vieweg2021, Vieweg2023a} -- but not the decreased $\Nuexp$. 
Questioning the experimentally present measurement procedure, we continued by resembling it in a controlled manner based on the exact high-resolution data from its digital twin and thus extended our previous work \citep{Vieweg2023} to examine the sensitivity to different experimental uncertainties. Although we find that (i) $4$ temperature probes are too few to correctly identify the mean upper solid-fluid interface temperature and (ii) the statistical heat transfer is clearly affected by the measurement procedure -- indicating a contradiction to our conclusion drawn in \citep{Vieweg2023} --, there is practically no impact of this procedure on the \textit{average} heat transfer.

Eventually, the comparison of the vertical velocities $u_{z}$ from both the experiment and its digital twin suggested a systematic underestimation in the case of the former. 
In fact, this out-of-plane velocity component is most susceptible to systematic measurement errors during stereoscopic PIV \citep{Westerweel1997, Prasad2000, Cierpka2012, Raffel2018} with an exact calibration being key \citep{Prasad2000, Wieneke2005}.
A subsequent re-assessment of the original stereo-PIV measurement data starting from the raw camera images revealed indeed a camera calibration error in \citep{Moller2021, Moller2022a, Moller2022} and thus also the data used in \citep{Vieweg2023, Teutsch2023}, underestimating the distances between the calibration planes. Correcting this mistake results in a $24 \%$ relative increase of $u_{z}$ -- the latter's sign and distribution, as well as the size and temporal evolution of flow structures, are not affected. This allows to finally to collapse the data from both the laboratory experiment as well as the numerical simulations across the entire range of considered $\Ra$, particularly with respect to $\Nu$. This resolves the remaining motivating discrepancies.

This study highlights that digital twins represent, together with the resemblance of laboratory measurement procedures, a highly useful tool for resolving discrepancies between experimental and numerical observations and thus drive the progress in thermofluid science with its numerous applications.
From an experimental perspective, our study suggests moving towards volumetric Lagrangian particle tracking techniques to prevent incorrect reconstructions of individual velocity components \citep{Kaufer2024}. Especially in combination with physics-informed machine learning, this allows revealing even more information of the flow \citep{Toscano2024}.
From a physical perspective, our study underlines the crucial role of realistic thermal boundary conditions with respect to the formation of long-living large-scale flow structures as well as their characteristic size and lifetime. It is clear that this point becomes even more important when the geometry of the heat transfer system goes beyond a simple cuboid configuration as discussed here. Understanding the effect of \textit{symmetric} non-ideal thermal boundary conditions is essential for a successful interpretation of more complex configurations and will be addressed in a future study. Another point that has to be left open for future work is the impact of these non-ideal boundary effects at higher Rayleigh numbers.

\begin{Backmatter}
\paragraph{Acknowledgement}
The authors are very grateful to Bernd Wieneke and Dirk Michaelis from LaVison GmbH for discussing the impact of the different camera calibration models.

\paragraph{Funding Statement}
P.P.V. is funded by the Deutsche Forschungsgemeinschaft (DFG, German Research Foundation) within Walter Benjamin Programme 532721742. 
T.K. and C.C. are supported by the Carl Zeiss Foundation within project number P2018-02-001 \enquote{Deep Turb – Deep Learning} and by the DFG within project number 467227170.
P.P.V. gratefully acknowledges the computing centre of the Technische Universität Ilmenau for providing access to, as well as computing and storage resources on its compute cluster MaPaCC24.

\paragraph{Declaration of interests}
The authors report no conflict of interest.

\paragraph{Author Contributions}
P.P.V. designed the study, prepared and performed the numerical simulations, plotted the figures, and drafted the manuscript.
T.K. provided the experiment data based on \citep{Moller2021, Moller2022} and re-processed it with C.C..
All authors contributed equally in discussing the data and finalising the paper.

\paragraph{Data Availability Statement}
Supporting data to this study can be made available by the authors upon reasonable request.

\bibliographystyle{jfm}

\end{Backmatter}
\end{document}